# GRAPHICAL ABSTRACT & HIGHLIGHTS

- Very concentrated cohesive suspensions show non-monotonic flow curves.

- This behaviour is a consequence strain-rate softening of the particle phase stress.

- We demonstrate how to characterize difficult suspensions with seemingly irreproducible flow properties.

A non-monotonic flow curve for a coagulated $CaCO_3$ suspension.

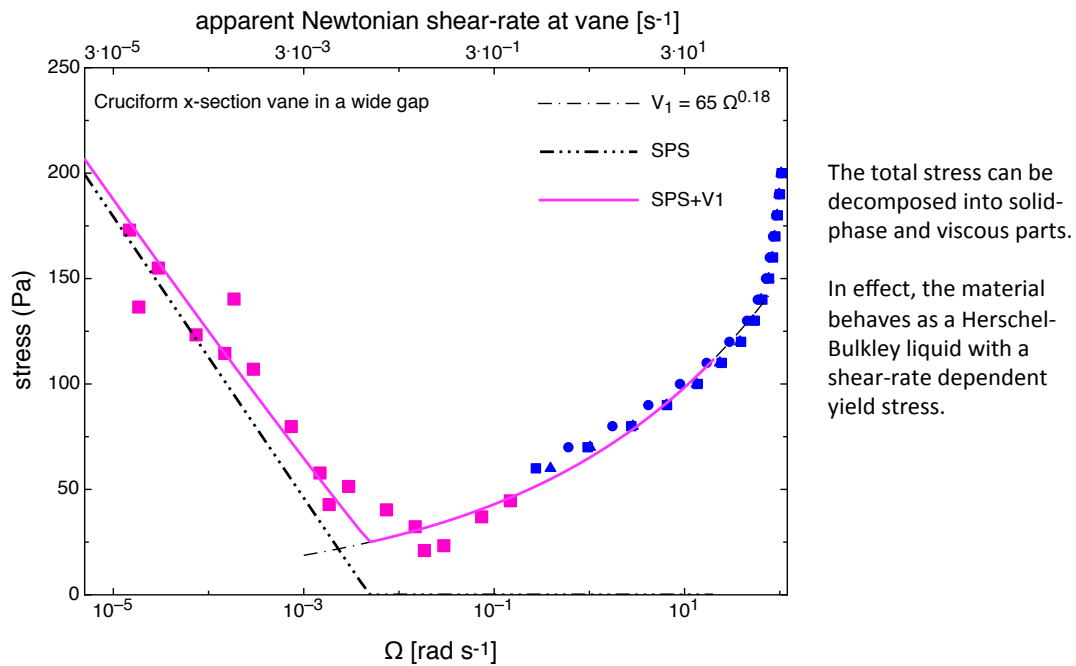

The total stress can be decomposed into solid-phase and viscous parts.

In effect, the material behaves as a Herschel-Bulkley liquid with a shear-rate dependent yield stress.

# The non-monotonic shear-thinning flow of two strongly cohesive concentrated suspensions.

Richard Buscall[1,2,3][*], Tiara E Kusuma[2], Anthony D Stickland[2], Sayuri Rubasingha[2], Peter J Scales[2], Hui-En Teo[2] and Graham L Worrall[3]

[1] MSACT Research & Consulting, 34 Maritime Court, Exeter, EX2 8GP, UK.

[2] Particulate Fluids Processing Centre, Dept of Chemical & Biomolecular Engineering, University of Melbourne, Parkville, Vic. 3010, Australia.

[3] ICI Research & Technology, The Wilton Centre, Redcar, TS10 4RF, UK.

*Author to whom correspondence should be addressed, email:r.buscall@physics.org.

Dedicated to Professor Ken Walters FRS in the year of his 80th Birthday.

## Abstract

*The behaviour in simple shear of two concentrated and strongly cohesive mineral suspensions showing highly non-monotonic flow curves is described. Two rheometric test modes were employed, controlled stress and controlled shear-rate. In controlled stress mode the materials showed runaway flow above a yield stress, which, for one of the suspensions, varied substantially in value and seemingly at random from one run to the next, such that the up flow-curve appeared to be quite irreproducible.  The down-curve was not though, as neither was the curve obtained in controlled rate mode, which turned out to be triple-valued in the region where runaway flow was seen in controlled rising stress. For this first suspension, the total stress could be decomposed into three parts to a good approximation: a viscous component proportional to a plastic viscosity, a constant isostatic contribution, and a third shear-rate dependent contribution associated with the particulate network which decreased with increasing shear-rate raised to the -7/10th power. In the case of the second suspension, the stress could be decomposed along similar lines, although the strain-rate softening of the solid-phase stress was found to be logarithmic and the irreducible isostatic stress was small. The flow curves are discussed in the light of*

*recent simulations and they conform to a very simple but general rule for non-monotonic behaviour in cohesive suspensions and emulsions, namely that it is caused by strain-rate softening of the solid phase stress.*

## 1. Introduction

It is generally accepted that for suspensions to show a yield stress one of two circumstances has to apply: either the particles need to be crowded such that their zones of influence (e.g. ion-atmospheres) overlap, or they need to attract each other sufficiently strongly for a ramified network having the properties of a viscoelastic solid to form. The paper is concerned with the yield and flow of two very concentrated and strongly cohesive suspensions which showed rather complex and variable behaviour inasmuch that in controlled-stress testing they showed erratic yield and flow hysteresis, whereas in controlled-rate they showed a rate-dependent yield stress, resulting in a non-monotonic flow curve.

Not all cohesive suspensions with a yield stress show reproducible and history-independent flow behaviour, even if some, e.g. flocculated clay dispersions, do [1]. Certain others, including some coagulated mineral suspensions and coagulated latexes can be changed irreversibly by subjecting them to sustained shear flow [2-4]. Indeed, in suspension rheometry, a rather common ploy has been to subject samples to high-shear pre-conditioning in order to tame them and obtain reproducible results thereafter [2-4]. Furthermore, it has been found that although shearing at controlled rate can have significant irreversible effects on the structure in some cases, the effects of shearing with control of stress can be much more profound, as was demonstrated by Mills et al. [4], who showed that ramified particulate gels could be converted into suspensions of uniform dense granules by means of the application of controlled-stress conditioning. The problem with pre-conditioning as an aid to rheometry though is that one never then knows what has been thrown away in terms of the native behaviour of the material. There are reasons then to suspect that the published literature on the flow of cohesive suspensions could well be biased towards systems that show reproducible behaviour intrinsically (ergodic systems), or which can be taken to a local stable state by some pre-conditioning procedure.

Here we present representative flow curves for two materials that appeared at first sight to be far from well-behaved and which in controlled stress testing in particular, showed erratic yielding inasmuch that the yield stress seemed to vary substantially but at random from one run to the next. The use of alternative test protocols showed however that the irreproducibility seen in controlled stress was a consequence of the materials having a highly non-monotonic flow curve.

Non-monotonic flow curves have been reported for a wide range of complex fluids, including suspensions and emulsions [5,6]. It is now well-understood that non-monotonic flow curves lead to steady-state, or, "stable" shear-banding, hence the two problems tend to be explored together [5-11], even if shear banding is seen to be a more general phenomenon when the possibility of temporary or transient banding is admitted [12].

For flow curves to show a maximum and minimum there has to be shear-thinning of specific kind: it is not enough that the material yield and shear-thin in the usual sense, since that alone can only give monotonic curves of the type captured by the Herschel-Bulkley model [13]. Rather, the stress transmitted through the particulate network has to decrease with increasing strain-rate [7]; or, to put it another way, there has to be progressive strain-rate softening of the elastic stress. At an operational level then, there seems to be very little mystery since all one needs to do to generate non-monotonic flow curves schematically is to modify the yield term in, say, the Herschel-Bulkley equation by factoring in a suitable strain-rate thinning function, $g(\dot{\gamma})$, thus, $\sigma(\dot{\gamma}) = \sigma_y g(\dot{\gamma}) + k\dot{\gamma}^n$. The interplay of the yield and viscous terms will then generate non-monotonicity, given that the strain-thinning function is strong enough and the yield stress large enough (see Appendix 1 for more detail). The real questions then, concern the origins and nature of the strain-rate softening of the solid phase elastic stress. The current state of knowledge suggests that the suspension should probably be very concentrated and also, perhaps, cohesive enough to be non-ergodic [7,14-17].

Of particular interest in the present context are recent simulations of the flow of a large assemblage of concentrated athermal cohesive discs sheared between rough boundaries (adhesive yield or slip is not just an experimental problem) performed by Irani et al. [17] using LAMMPS [18]. Earlier work [19] on a very similar but over-crowded system (the Lennard-Jones type inter-disc potential used allows over-crowding) found transient shear banding only, whereas when the concentration was dropped below the jamming value stable shear-banding was seen [17]. The high shear or "repulsion dominated", branch of the flow curves showed Bagnold grain-inertia scaling [19, 20] and, remarkably, the low-shear or "attraction-dominated", region of the flow curves where banding occurred could be scaled simply too [17]. A key quantity in the scaling was the local co-ordination number relative to the minimum value for isostatic transmission of stress. The non-monoticity and banding seen in the simulations was found to be a consequence of the non-ergodicity. More specifically, it could be attributed to asymmetry in the dynamics of structural change: in these athermal systems the dynamics are controlled by shear-rate alone and the nature of the dependence of local structure on shear-rate was key. In crude terms, the simulations generated the right kind of strain-rate thinning. The scalings found in [17] suggest that the control parameters for cohesive particles are the distance from jamming in terms of concentration and the strength of attraction, together with the characteristic time for recovery. That the latter matters could be taken to imply that whereas the system should be non-ergodic, it need not be athermal necessarily.

The first of our two suspensions, a concentrated and heterocoagulated mixed pigment suspension, was a chance find. Later, and as a preliminary test of the idea that any strongly cohesive suspension might show non-monotonic behaviour if concentrated enough, a second suspension was prepared comprising coagulated calcium carbonate particles at a volume fraction of 0.4. This suspension was found to show the same kind pronounced non-monotonic behaviour as the first, albeit with quantitative differences.

## 3. Experimental details

### 3.1 Materials

The first material used was an industrial intermediate in the form of heterocoagulated pigment mixture based on rutile (titania) of ca. 300nm particle diameter mixed with a minor component of significantly smaller size. It was chosen for rheological investigation in the first instance simply because it showed pronounced instability in processing flows. It was subsequently found to show spectacularly non-monotonic flow in rheometric testing, this being the cause of the process flow instabilities, presumably. We had seen flow curves with modest maxima before, as perhaps have most who have worked on concentrated cohesive suspensions, but nothing nearly so pronounced as this. The volume fraction of solids was ca. 0.45 notionally, although optical and electron microscopy showed the primary particles were agglomerated into secondary framboid clusters of up to a few μm in diameter; hence the effective volume-fraction was higher, as it can in aggregated systems. The high-shear relative viscosity was found to be of order 500 (see fig. 2 below) hence the secondary particles were very close to being jammed in effect. We found that the rheology was independent of shear history, even though this is by no means always true for coagulated and flocculated systems [2-4]. It could however just have been perhaps that the processing history of the material as received had already brought it to a stable state. In summary then, the first material was a concentrated suspension of polydisperse framboids with sizes in the 1 to 10 μm range typically. It had a high plastic relative viscosity, implying that the framboids were close to being jammed.

The second material was commercial calcium carbonate (Omyacarb® 2-LU, Omya California Inc.) suspended in 0.01M of potassium nitrate and coagulated at the natural pH of 8.2 +/- 0.5. The weight-average mean particle size measured using a Malvern Mastersizer 2000 was 4.5 μm. The volume-fraction was 0.4, based on the manufacturers figure for the density of the particles of 2700 kg m$^{-3}$. The differential high-shear relative viscosity was of order 30, which again implies an effective volume-fraction much higher than the actual of 0.4.

## 3.2 Methods

In all, five different rheometers were employed, two in one laboratory for the pigment suspension and three in another for the CaCO$_3$ suspension. In the case of the pigment suspension, the rheological measurements in concentric cylinder flow were made either in controlled-stress mode using a Bohlin CSR rheometer or in controlled-rate mode using a Haake RV1 instrument. The measurements were made using either a roughened (milled) 14 mm dia. cylindrical bob in a 25 mm internal dia. cup, or, mostly, with a 14 mm diameter cruciform cross-section vane in a 25 mm internal dia.

cup, with very similar results being obtained with each tool. In stress control, the stress applied was increased in equal steps starting well below the yield point. In controlled-rate the angular velocity was increased in logarithmically spaced steps starting from a low value of ca. 100 µrad s$^{-1}$. The dwell time at each step was typically 120s, comprising 60s "equilibration" and 60s signal-averaging. Strictly-speaking, it would be preferable to choose the dwell time in order to ensure that a steady-state was reached at each step, except that the times then need to be very long at low shear rates and stresses for materials like this. The problem then is prevention of water uptake or loss: water-traps will prevent loss but tend to produce water-uptake, albeit very slight. The latter is not usually a problem in general but it becomes one very near jamming where any change in water content, no matter how slight, can have a discernable effect. There was then a concern to keep run times to a minimum and samples were refreshed frequently.

The measurements on the CaCO$_3$ suspension were made in controlled-stress mode using a TA Instruments AR-G2 rheometer and in controlled-rate mode using a Rheometric Scientific$^{TM}$ ARES rheometer and a Haake VT550 instrument. It was not possible to use the same vane and cup in each instrument because the couplings were incompatible. The vane and cup dimensions used are given in Table 1:

**Table 1:**
Dimensions of test geometries for AR-G2, ARES and Haake testing
of the CaCO$_3$ suspension.

|  | AR-G2 | ARES | Haake |
|---|---|---|---|
| **Vane Diameter (mm)** | 28 | 8 | 25 |
| **Cup Diameter (mm)** | 142 | 34 | 75 |
| **Cup to Vane Diameter Ratio** | 5.07:1 | 4.25:1 | 3:1 |
| **Vane length (mm)** | 42 | 16 | 50 |

In controlled stress, stepped stress experiments were performed from 0 to 200 Pa in increments of 10 Pa held at 20, 40 and 60 seconds at each step. Controlled rate testing involved measuring the steady-state stress for a series of logarithmically spaced angular velocities covering several decades. In between each run, the suspension was homogenised by mixing for 1 minute and tapped lightly to release bubbles for 30 seconds. The vane was then lowered gently into the suspension and an equilibration time of 5 minutes was observed for all experiments before the tests were initiated. Water traps were used to alleviate moisture loss from the system: because the effective volume fraction calculated from the high shear viscosity was somewhat lower than the pigment suspension, the possibility of very slight water uptake was less of a concern.

## 4. Results & Discussion

Fig. 1 shows flow curves for the pigment suspension plotted on a semi-log scale in the form of shear stress versus the logarithm of the angular velocity.

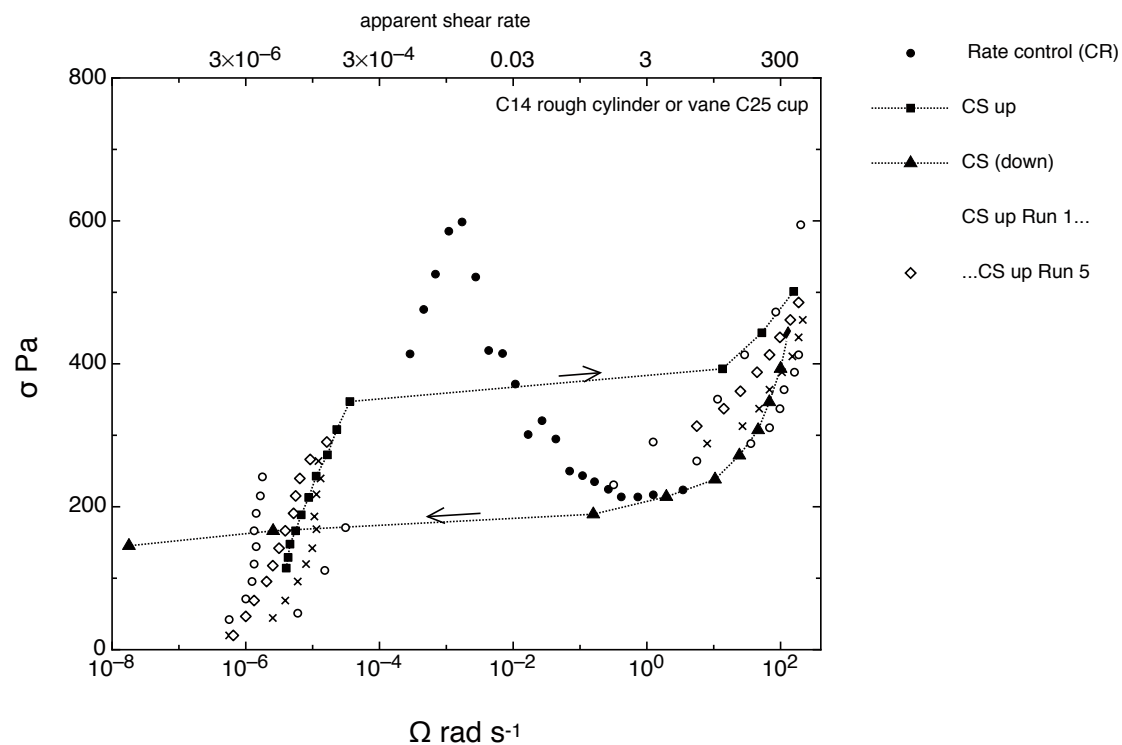

**Fig. 1.** *Flow curves plotted as shear stress versus the angular velocity of the 14 mm diameter rotor. The corresponding, apparent Newtonian shear-rate is given on the upper ordinate for guidance purposes. Filled circles – controlled rotation rate. Connected filled squares – ascending staircase of controlled stress; connected filled triangles – descending. Remaining points – further examples of many runs in ascending controlled stress mode.*

Because of the wide gap used and the highly non-linear behaviour, it was not possible to convert the latter into shear-rate, except very approximately. Nevertheless, the apparent Newtonian mean shear-rate is shown on the upper axis for guidance regarding order of magnitude (the notional shear-rate being a little less than three-times the angular viscosity in this case). Data is shown for two batches of material, BX4 and BX5, made on different days, although they had indistinguishable properties to all intents and purposes, as did several other batches investigated.

Three types of test were performed: (i) a staircase of controlled stresses upwards, (ii) like-wise, but followed by a descent, and (iii) a staircase of controlled rotation-rates upwards. The data shown are a subset of that taken overall, but representative. The reader is asked to focus on the unconnected points at the lower left in the first instance, these corresponding to ascending stress runs. An unsteady mean rotation rate of order $1 \mu \text{rad s}^{-1}$ was recorded typically, the dwell-times used being insufficient for retarded-elastic equilibrium to be reached below the yield stress, hence the small but non-zero values plotted should not be taken to imply viscous flow, but rather just viscoelastic deformation below the yield-point. At some point the angular velocity jumped by several orders of magnitude to a value between ca. 0.1 and 1, or thereabouts. The rotation rates prior to the jump and the jump stress, or, apparent yield stress, varied from one run to the next, seemingly at random; the six sets of data shown for six aliquots from two batches of material being a small subset of a very large number of runs performed overall, most of which data is not shown for the sake of clarity. Overall, the run-to-run variation of the jump or yield stress ca. 250%.

By turning attention to the RHS of the diagram it can be seen that the high shear flow behaviour was variable too, even though the individual runs seem to converge at the highest rates of $100 \text{ rad s}^{-1}$ or so. Higher rates could not be reached because the upper speed cut-out of the rheometer would then terminate the run.

The connected points in fig.1 show the result of returning back down the staircase of stresses. Substantial hysteresis between ascent and descent is evident, although further runs showed that, unlike the up-curve, the down-curve was reasonably reproducible. The return stress of ca. 150 Pa was found to coincide approximately with the lowest yield stress recorded in ascent mode; the implication being then that the material could yield anywhere above this value, seemingly at random.

The filled circles lying between rotation rates of ca. $10^{-4}$ and $5 \text{ rad s}^{-1}$ were obtained by testing in controlled rotation rate mode (CR) and these revealed the triple-valued nature of the flow curve in the region inaccessible to controlled rising stress. Again, this curve was acceptably reproducible and showed no hysteresis upon ascent and descent. The controlled rate measurements imply a yield stress of ca. 600 Pa, ca. four times the minimum yield stress seen in controlled stress (CS), and significantly larger than the upper value seen in CS, this being ca. 420 Pa. That is not to say that a value as large 600 Pa might not have been seen in controlled stress ascent eventually, had we persisted further perhaps. Be that as it may, it can be seen also that a stress of 600 Pa was not reached again until flow rates ca. 100,000 times higher were achieved.

Flow curves with hysteresis are far from new of course, one well-established cause being thixotropy, and another being intrinsic non-monoticity of the type shown [7]. The curves shown are however rather spectacular examples in terms of the scale of the effect. We were caused to recall that we had variously seen materials showing either seemingly irreproducible flow curves or humps, albeit small ones, many times before, but nothing on anything like this scale.

The true or underlying steady flow-curve suggested by the data then, follows the controlled rate points and then continues up the path of the descending curve in controlled stress. This curve is re-plotted figs. 2 and 3 on linear and logarithmic scales. The high shear plot in fig. 2 is linear above 25 rad s$^{-1}$ with a slope of 1.6 Pa.s rad$^{-1}$, implying a constant plastic viscosity. A lower bound to the value of the latter can be obtained from the Newtonian apparent shear-rate conversion factor, to give a value of 0.55 Pa.s. This in turn implies a plastic relative viscosity $> 550$, suggesting that the effective volume-fraction was high: it insertion into the Krieger-Dougherty equation [22], say, would imply $> 0.98\varphi_{max}$. Notice too from fig. 2 that had we confined our testing to controlled rate at rotation rates of above 1 rad s$^{-1}$ only, then nothing outlandish would have been detected, we would just have seen a Bingham plastic. In controlled rate testing, the complexity is only revealed at low rates, whereas in controlled stress mode it just cannot be avoided.

The unfilled diamonds in fig. 3 show the total stress with the viscous contribution calculated from the plastic viscosity subtracted. It can be seen that this asymptotes to a more or less constant value of 215 +/-10 Pa above 1 rad s$^{-1}$. Subtracting this plateau stress in turn gives the unfilled squares, which, above the initial strain-controlled rise, decay with strain-rate as a power law of exponent ca. -0.7. The data become very sensitive to the precise value of the plateau stress used as the plateau is approached of course, as shown by the open circles, and hence the last points shown were ignored in the fit. It is evident then that the whole curve can be modelled or fitted using strain and strain-rate softening functions together with a high shear viscosity term, provided, that is, that it is assumed that there are two parts to the elastic stress: a mutable part degraded by shear rate and an immutable or persistent part. The basic ideas are consistent with what was seen in the 2-d simulations of Irani et al. [17] in so far as they go. The detail is very different though since very much weaker logarithmic strain-rate softening was seen in the simulations.

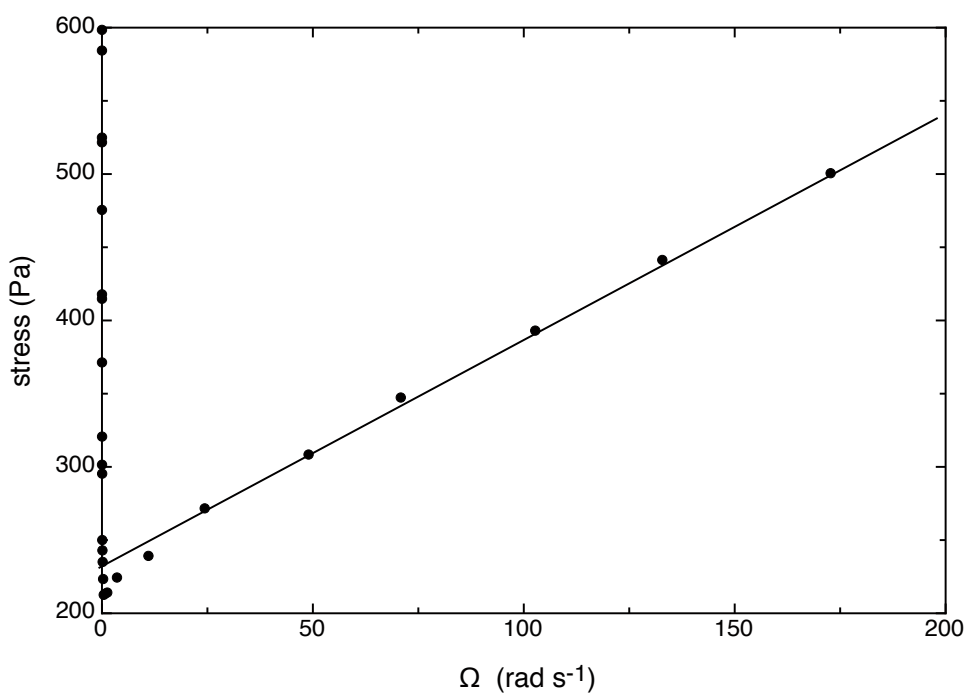

**Fig. 2.** *The multi-valued flow curve suggested by fig. 1 plotted on a linear scale. The drawn line has a slope of 1.6 Pa.s rad$^{-1}$, which equates to an apparent plastic viscosity of ca. 0.55 Pa.s, based on the apparent Newtonian shear-rate, and hence a relative viscosity of > 500.*

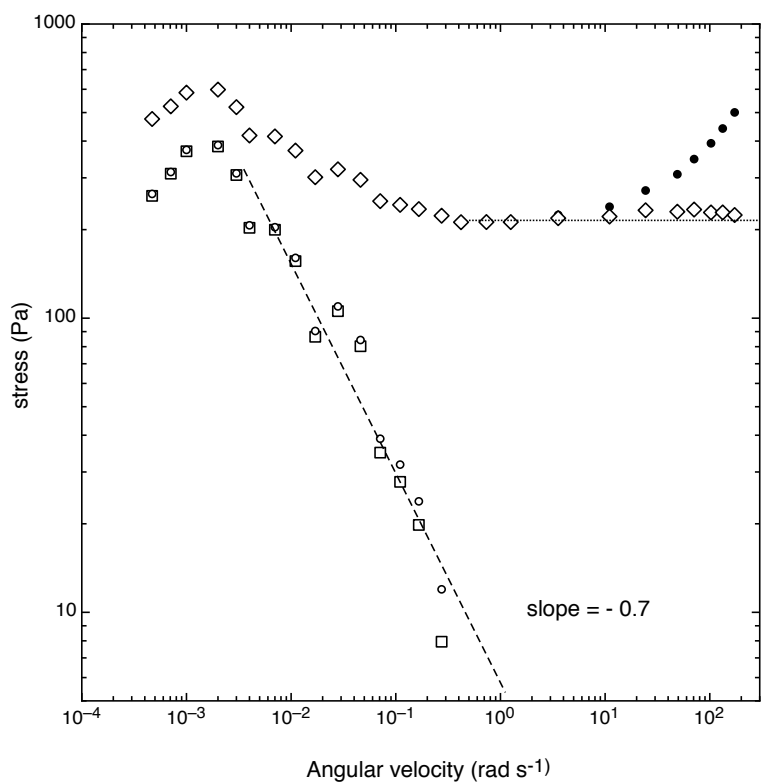

**Fig. 3** - *Filled circles - the multi-valued flow curve suggested by fig. 1. Unfilled diamonds - the same data but with the viscous stress calculated using the plastic viscosity obtained from the linear plot show in fig.2. Unfilled squares and circles with the plateau stress subtracted, using values of 215 and 211 Pa, respectively.*

In order to test the idea that a combination of supra-colloidal particle size, high volume-fraction and strong cohesion suffice to generate non-monotonic behaviour, similar measurements were performed on a better-defined suspension of coagulated $CaCO_3$ particles at a volume fraction of 0.40. Three rheometers were employed overall, two for the controlled strain-rate measurements and one for the controlled stress. A cruciform cross-section vane tool was used in each case, albeit a different one in each case because of incompatibility of the rheometer couplings. Because of this the data are plotted against the product of angular velocity and vane radius in order to scale the results when comparisons of controlled stress and controlled strain are made below. The rational for this was that the shear-rate at the cylindrical surface swept by the vane would be in strict proportion to this ratio for a vane in an infinite sea and thus approximately so for a shear-thinning liquid in a wide gap. First, though, fig. 4 shows the controlled stress data plotted on a linear scale.

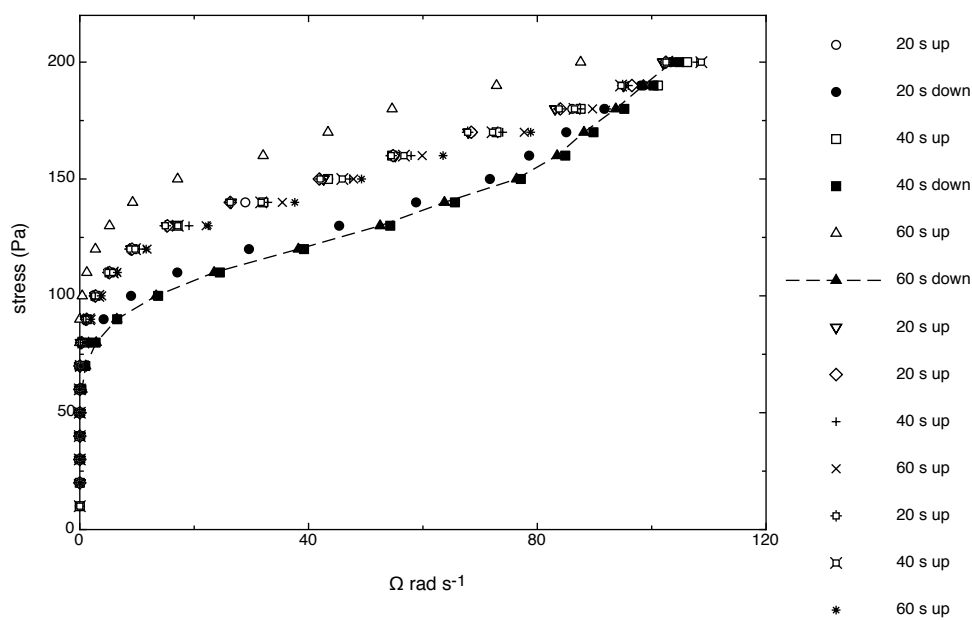

**Fig. 4** - *Flow curves from controlled stress for three different dwell times per point. The filled points denote the down-curves. The hysteresis was not overtly time-dependent and thus not a manifestation of thixtropy in the usual sense.*

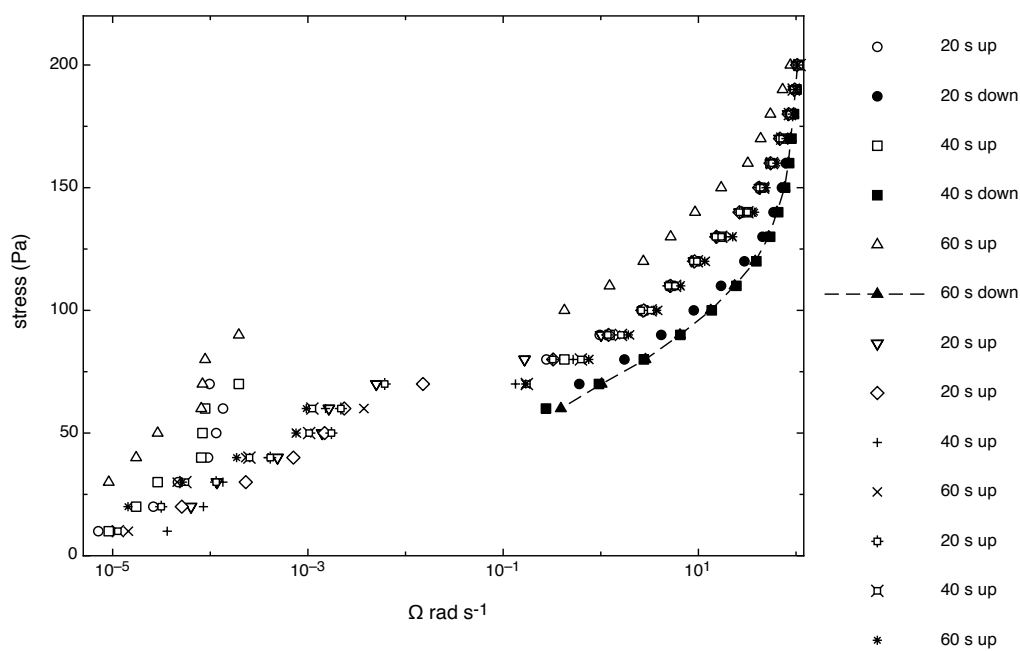

**Fig. 5** - *As fig. 4 but semi-log. The yield behaviour was erratic; there was no systematic effect of dwell time.*

Three different dwell times were used at each stress point. The shear thickening at high rate apart, any one of the curves looks somewhat like a Bingham plastic when plotted on a linear scale. The down-curves were more reproducible, just as they were with the pigment suspension. The semi-log plot, fig. 5, shows that the yield was similarly erratic, even if the variation was not quite so large as it was for the pigment suspension.

Controlled stress and controlled strain are compared in figs. 6 and 7, which shows also the peak stress in controlled strain, discussion of which is relegated to Appendix 2. Overall, the pattern of behaviour is broadly similar to that of the pigment suspension, qualitatively speaking, there are however quantitative differences: the stresses are ca. 1/3 of those seen for the pigment suspension overall and the minimum stress is lower still, relatively-speaking. Furthermore the high shear branch shows first viscous shear thinning followed by shear thickening and the shear-rate softening of the solid-phase stress now looks to be more logarithmic than power-law. The viscous branch can be mimicked quite well below the shear-thickening by a power-law with an exponent of ca. 0.18, although, in view of the scatter, a Herschel-Bulkley model with a small yield stress (< 5 Pa) would probably work equally well. It is clear though that the steady-state flow-curve implied by the data can again be decomposed into a shear-rate softening solid-phase contribution plus a viscous term. There is however no need, necessarily, for a "plateau stress" term in this case, given that the minimum stress is small, which implies that the remnant isostatic stress is close to zero. In this case the yield stress obtained by extrapolation from fig. 4 of ca. 80Pa is very much larger that the remnant isostatic stress (of < 5Pa) because of the shear thinning of the viscous branch. Another way to say this is to note that that Bingham model and Herschel-Bulkley extrapolations, both of which can look sensible, depending upon how the data are plotted, would give very different values for the yield stress. Since the high shear viscosity did not approach a constant in this case, the Krieger-Dougherty equation was applied to the minimum value of the apparent *differential* viscosity of ca. 30 to obtain a lower estimate of the effective volume fraction of ca. $0.90\varphi_{max}$ (cf. $0.98\varphi_{max}$ for the pigment suspension).

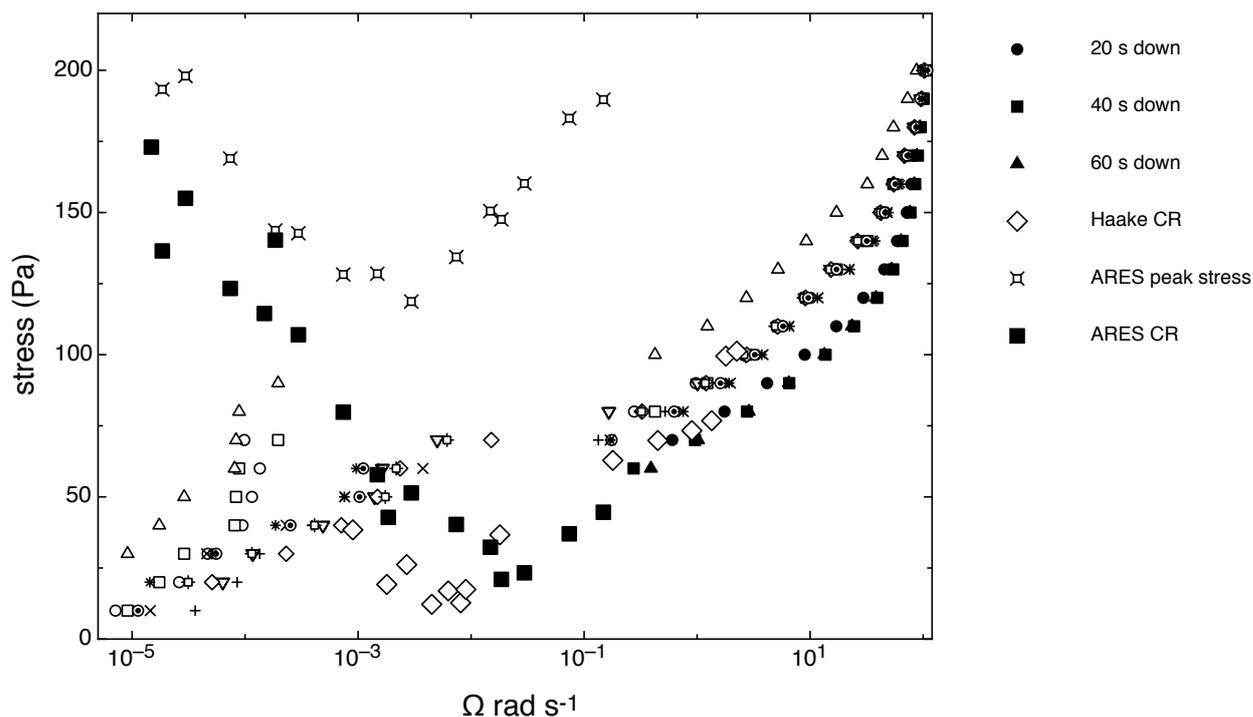

**Fig. 6** - *Comparison of controlled stress with steady-state data from controlled-rate measurements. Also shown is the peak stress measured with ARES during flow-start-up. The vane diameters were 8.1mm (ARES), 24mm (Haake) and 28mm (AR-G2) and to compensate for this the rates for the ARES and the Haake have been scaled by factors of 8.1/28 and 24/28, respectively.*

In some respects the data for the $CaCO_3$ suspension are somewhat more akin to the simulations of Irani et al. [17] than was that for the pigment suspension, inasmuch that the strain-rate softening of the solid-phase stress looks logarithmic and the viscous branch shear-thinning. It should be mentioned though that the shear thinning in Irani et al. was granular in origin, as the simulations were liquid-free. It has however been demonstrated in some intriguing simulations very recently that very concentrated model suspensions can look granular when the liquid phase viscosity is low [30].

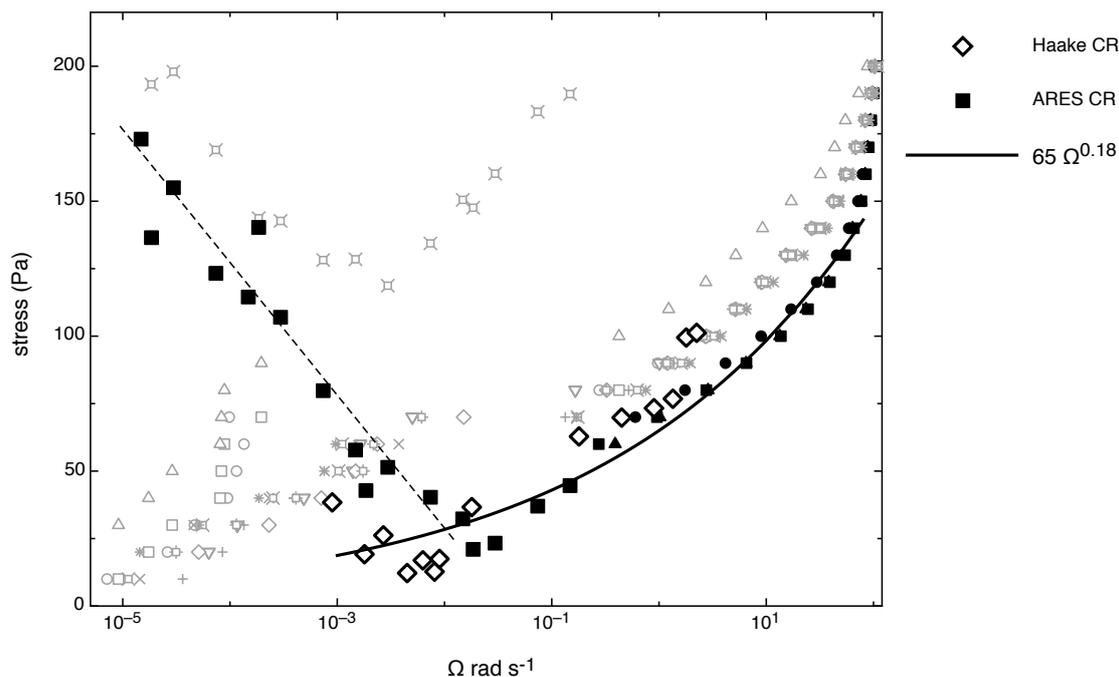

**Fig. 7** - *As fig. 6 but with the CS up curves and the CR peak stress points greyed out for clarity. The higher shear branch is compared with a power-law up to the point where shear thickening sets in.*

We will conclude the discussion by suggesting some possible directions for a future study of a more systematic kind.

(i) The effect of dilution and the effect of the viscosity of the liquid phase: it has been suggested that whereas strain-rate softening might be a necessary condition for non-monoticity, the solid-phase stress needs to dominate the viscous stress too, hence the idea here would be to try and change the balance between the two and observe the effect.

(ii) Pressure-driven test flows in one and two dimensions looking for flow instabilities (these were seen when processing the titania pigment system but were not quantified).

(iii) Effects of particle size and, perhaps, particle roughness: the latter might be assessed by comparing, say, framboids made using the method of Mills et al. [4] (which works more widely and not just on the emulsifier-free polystyrene latex used by them [R Buscall & J I McGowan – unpublished work]), with, say, Ballotini or calcium carbonate of similar size.

(iv) Effect of interaction strength, the variation of which could be done using fatty-acid stabilised alumina or $CaCO_3$ particles in oil following Bergstrom [27], by which means the attractive well-depth can varied by an order of magnitude or more.

## Conclusions

A strongly cohesive, non-ergodic and highly concentrated pigment suspension was found to exhibit a highly non-monotonic flow curve in controlled-rate testing and erratic yield in controlled stress testing. Both features were found to be a result of pronounced shear-rate softening of the non-Newtonian or particulate phase stress over a range of strain-rates where the viscous stress was small. The strain-rate softening appeared to follow a power-law with an exponent of -0.7, initially, and then to plateau, implying that there was an irreducible level of isostatic stress. The behaviour seen was somewhat similar qualitatively to that found in simulations of sticky athermal discs near the jamming transition performed by Irani et al. [17], albeit that the strain-rate softening was very much stronger in the experiments. The simulations of Irani et al. [17] could be taken to suggest that any strongly cohesive system of spheres near jamming might or could show similar behaviour and it was indeed found that a 40%v/v suspension of strongly-flocculated 4μm $CaCO_3$ particles showed the same sort of behaviour as the pigment, qualitatively, although the strain-rate softening of the solid-phase was more progressive and the remnant isostatic stress was close to zero, the effective volume-fraction being somewhat lower than in pigment suspension (90% of jamming as opposed to 98%, as deduced from the minimum high-shear viscosity using the Krieger-Dougherty equation [22]). On the other hand, the $CaCO_3$ suspension showed shear thickening at the highest shear-rates too, which the pigment suspension did not; because it was more polydisperse perhaps. It is argued (Appendix 1) that strain-rate softening of the particulate phase stress is a necessary condition for non-monotonic stress. In appendix 2 it was shown that that the dynamics of stress growth and decay for the $CaCO_3$ suspension were strain-rate controlled, as is to be expected where the combination of particle size plus strong cohesion render Brownian motion unimportant.

It might be appropriate perhaps to attempt provide a somewhat more prosaic or practical summary too, thus: suspensions like those tested present in different ways depending upon how they are tested, or, more generally, how they are caused to flow. They can look quite simple in controlled-rate flow at higher rates and show Bingham or Herschel-Bulkley flow. At low rates, rates readily achieved in modern rheometers, but unlikely to be encountered in larger scale kinematically controlled flows, they can look more complex and show a rate-dependent yield stress, in effect. They look their worst in controlled stress and hence, by implication, pressure-controlled flows, where yield is erratic and where shear banding will occur. We suspect fluids like the two we have studied are probably far from being uncommon. Rather, we suspect that such behaviour can easily be missed in routine flow-curve testing at controlled rates and readily dismissed as "irreproducible" in controlled stress.

**Appendix 1 - Yield and strain versus strain-rate softening of the solid phase stress.**

The notion of a yield stress has utility in 1-dimensional simple shear flow, even though there can actually be no such material property as such, given that the true yield criterion has necessarily to be an invariant of the stress tensor more generally, and an invariant common to both stress and strain, arguably. An obvious candidate for such an invariant is the stored, or elastic, strain energy, this being compatible or consistent with the successful von Mises yield criterion [23]. The strain energy criterion for yield does not of necessity imply a sharply-defined yield stress in simple shear flow, since the viscoelasticity below the yield point can lead to a yielding over a range of stresses. The simplest case though would be where the material is simply linear-elastic, as opposed to viscoelastic, up to a critical strain, which then does imply a definite yield stress. The presence of retarded elasticity, on the other hand, suffices to explain time-dependent yield in creep and a yield stress range [R Buscall et al., paper in preparation], just as is seen widely seen in practice [24-26], even though it is often misunderstood to be a manifestation of thixotropy [26]. The width of the yield stress range is predicted to depend upon the ratio of the retarded to instantaneous compliance. Having said that, it will however suffice here to consider an imaginary model material that is purely non-linear elastic up to the yield-point. The Herschel-Bulkley model for steady flow can then be written as,

$$\sigma = \sigma_y + k\dot{\gamma}^n \quad \rightarrow \quad G_0\dot{\gamma}\int h(\dot{\gamma}t')dt' + k\dot{\gamma}^n \quad ;$$

where $h(\gamma)$ is some differential strain-softening function which has the property that it goes to zero above a certain strain, the simplest case of which would be a step function such that $h(\gamma) = 1$ below some critical strain and zero above, giving a yield stress of $\sigma_y = \gamma_c G_0$. The HB equation on the LHS gives a monotone increasing stress with shear-rate of course, and so does its elaboration on the RHS, regardless of the nature of the strain-softening function chosen. One can generalise further and make the material non-linear *viscoelastic* below the yield point, except that the conclusion remains the same; viz. that *strain-dependent softening (or more generally strain-energy softening) alone can only give monotonic flow curves*.

The same is not true if *strain-rate softening* (or more generally, perhaps, dissipation-rate softening ) of the solid-phase stress is allowed. For controlled shear-rate flow, the strain and strain-rate dependencies have necessarily to be factorisable in a steady-state and so any strain-rate softening of the solid phase stress can be accounted for formally simply by multiplying the first term in the equation above by a strain-rate dependent softening function, $g(\dot{\gamma})$ say, to give $\sigma = \sigma_y g(\dot{\gamma}) + k \dot{\gamma}^n$. It is then easy to see that the resulting flow curve can be non-monotonic if the first term decreases with increasing shear-rate and it does so before the viscous term becomes too dominant. One obvious constraint on the softening function $g(\dot{\gamma})$ is that the energy dissipation $\sigma \dot{\gamma}$ rate should increase with strain-rate. Any strain-rate softening function with a decay weaker than $1/\dot{\gamma}$ is then guaranteed to give an increasing rate of overall energy dissipation with increasing shear-rate and so a constraint on the form of $g(\dot{\gamma})$ could be that it be weaker than that.

## Appendix 2 – Stress growth in step strain-rate.

The steady-state mean and peak stresses measured at controlled rate using the ARES were plotted in figs 6 and 7. The time-dependent stress growth is shown in fig. 8, the parameter being the apparent Newtonian shear-rate. The data are re-plotted against nominal strain, the product of time and apparent shear-rate, in fig. 9.

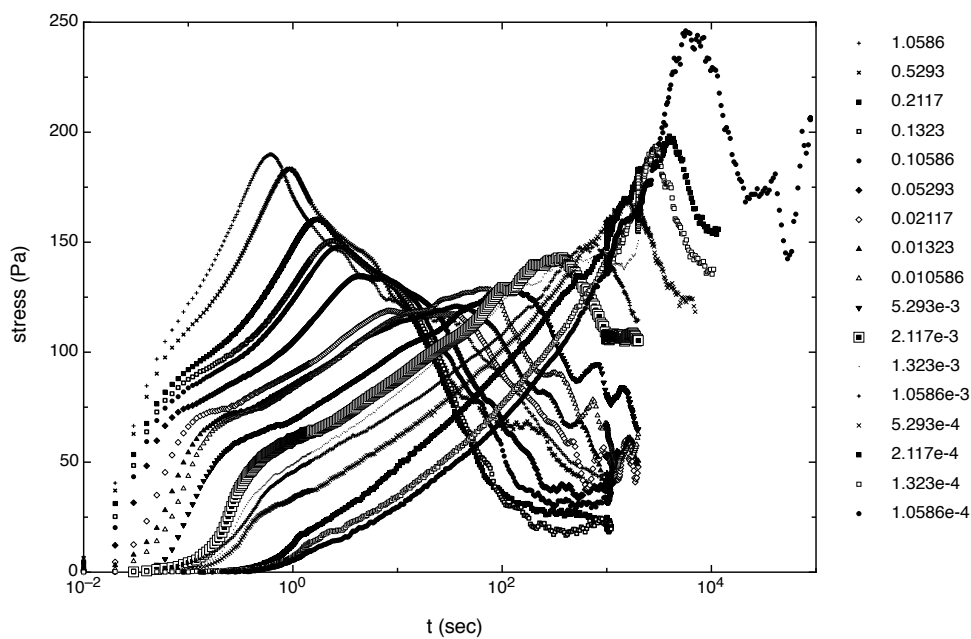

**Fig. 8** - *Stress growth and decay at controlled strain-rate using the ARES. The parameter is the apparent mean (Newtonian) shear-rate and proportional to the angular velocity.*

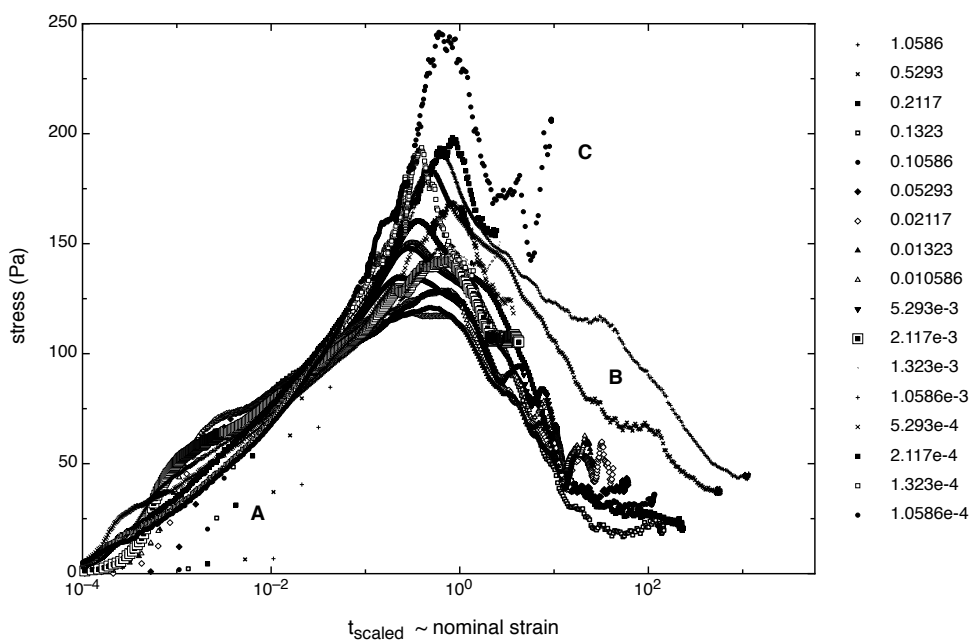

**Fig. 9** - *As fig. 8 but with time multiplied by the apparent strain-rate to give the engineering or nominal strain.*

The 4-decade or more range of time-scale is collapsed rather well onto curves of stress versus notional strain that peak at a strain of order unity [28, 29] and a Hencky strain of ~0.5. The outliers seen in region A are due to instrument inertia at the higher rates. The two overtly outlying curves in region B are those for the two highest rates, where the viscous stress is very significant.

The label C is meant to draw attention to fluctuations at long times for the lowest shear-rate, although the steady-state behaviour was always erratic to some extent on the solid-phase branch and increasingly so at as the rate was reduced, as if the system was hunting. Outliers not withstanding, the scaling could be taken to imply, perhaps, that the material has no intrinsic time-scale. The peak stress is located at a notional strain of order unity, a strain associated with "cage melting" according to Pham et al. [29].

The scaled rate of decay of the solid-phase stress is the approximately the same in each case, even for the two curves B since subtracting the viscous stress pulls these down to the others. The maximum stress is ca. 2.5 times as large as the extrapolated yield stress and ca. 2 to 3 times the yield stress seen in controlled ascending stress. The detailed interpretation and modelling of the transient behaviour will be taken up in a subsequent paper, as will more detailed modelling of the flow curves. The matter of estimating the true shear-rates and their distribution in the wide gap will also be taken up there.

**Acknowledgements:** Huntsman Pigments are thanked for permission to publish the data for the pigment suspension. Hui-En Teo was funded by a Brown Coal Innovation Australia Postgraduate Scholarship, Tiara Kusuma by a Melbourne International Research Scholarship of the University of Melbourne. Sayuri Rubasingha was funded by an Australian Postgraduate Award from the Australian Research Council. Infrastructure support at Melbourne was provided by the Particulate Fluids Processing Centre, a Special Research Centre of the Australian Research Council.